\documentclass[article]{elsarticle}

\usepackage{lineno,hyperref}
\modulolinenumbers[5]

\usepackage{amsmath,amssymb}
\usepackage{upgreek}
\usepackage{xcolor}
\usepackage{lineno}

\graphicspath{{fig/}}
\usepackage{multirow}

\journal{Journal of Instrumentation}

\begin{document}

\begin{frontmatter}

\title{The second production of RSD (AC-LGAD) at FBK}

\author[a]{M.~Mandurrino\corref{mycorrespondingauthor}}
\cortext[mycorrespondingauthor]{Corresponding author}
\ead{marco.mandurrino@to.infn.it}

\author[a,b]{R. Arcidiacono,}
\author[c,d]{A. Bisht,}
\author[c,e]{G. Borghi,}
\author[c,e]{M. Boscardin,}
\author[a]{N. Cartiglia,}
\author[c,e]{M. Centis Vignali,}
\author[d,e]{G.-F. Dalla Betta,}
\author[b]{M. Ferrero,}
\author[c,e]{F. Ficorella,}
\author[c,e]{O. Hammad Al\`i,}
\author[a]{A. D. Martinez Rojas,}
\author[a,f]{L. Menzio,}
\author[d,e]{L. Pancheri,}
\author[c,e]{G. Paternoster,}
\author[a,f]{F. Siviero,}
\author[a]{V. Sola,}
\author[a,f]{M. Tornago}

% The "\note" macro will give a warning: "Ignoring empty anchor..."
% you can safely ignore it.

\address[a]{INFN, Sezione di Torino,\\Via P. Giuria 1, 10125 Torino, Italy}
\address[b]{Universit\`a del Piemonte Orientale,\\Largo Donegani 2/3, 20100 Novara, Italy}
\address[c]{Fondazione Bruno Kessler,\\Via Sommarive 18, 38123 Povo (TN), Italy}
\address[d]{Universit\`a degli Studi di Trento,\\Via Sommarive 9, 38123 Povo (TN), Italy}
\address[e]{TIFPA-INFN,\\Via Sommarive 18, 38123, Povo (TN), Italy}
\address[f]{Universit\`a degli Studi di Torino,\\Via P. Giuria 1, 10125 Torino, Italy}

\begin{abstract}
In this contribution we describe the second run of RSD (Resistive AC-Coupled Silicon Detectors) designed at INFN Torino and produced by Fondazione Bruno Kessler (FBK), Trento.
RSD are \emph{n}-in-\emph{p} detectors intended for 4D particle tracking based on the LGAD technology that get rid of any segmentation implant in order to achieve the 100\% fill-factor. They are characterized by three key-elements, (\emph{i}) a continuous gain implant, (\emph{ii}) a resistive \emph{n}-cathode and (\emph{iii}) a dielectric coupling layer deposited on top, guaranteeing a good spatial reconstruction of the hit position while benefiting from the good timing properties of LGADs.
We will start from the very promising results of our RSD1 batch in terms of tracking performances and then we will move to the description of the design of the RSD2 run.
In particular, the principles driving the sensor design and the specific AC-electrode layout adopted to optimize the signal confinement will be addressed.
\end{abstract}


\end{frontmatter}

%\linenumbers

\section{Introduction}

One of the most crucial aspects in the 4D particle tracking with silicon detectors based on the LGAD (Low-Gain Avalanche Diode) technology is to obtain the maximum geometrical acceptance in a single plane of sensors. Such devices usually implement patterned $p$-type gain implants (the detector active area) surrounded by proper JTE (Junction Termination Extension) and $p$-stop structures, to avoid short-circuit between nearby pixels and to mitigate early breakdown phenomena. While, in reverse bias conditions, the gain layer provides the electric field responsible for the avalanche ionization of charges, in correspondence of the inter-pixel isolation implants we observe a reduced signal amplitude due to the absence of charge multiplication. This phenomenon produces a dead-area (or no-gain area) for particle detection, which deteriorates the almost hermetic coverage required by several of the most important high-luminosity experiments. To overcome this issue, we are currently developing the Resistive AC-Coupled Silicon Detector (RSD), a novel technology which gets rid of any segmentation structure and exploits a 100\% fill-factor (the ratio between the active and the total detector area) in the concurrent measurement of position and time of tracks. In the RSD design, the gain implant spreads over all the detector area and the spatial information of particle hits is reconstructed thanks to: (\textit{i}) a resistive $n^+$ cathode, with a given $R_\textrm{sheet}$ to allow the readout, and (\textit{iii}) a dielectric layer deposited on silicon, generating a capacitive coupling of such charges on the readout metal pads patterned on top of the dielectric.

By properly choosing the physical and technological parameters, it has been shown that the RSD readout scheme works in a wide range of pad pitch and size, down to very fine segmentation levels~\cite{2019Mandurrino_EDL}. Among these parameters we find $R_\textrm{sheet}$ ($\sim$k$\Omega$/$\Box$ ), directly depending on the $n^+$ dose, the dielectric composition and thickness, determining the coupling capacitance $C_\textrm{AC}$, and, finally, also the $p$-gain implantation dose, which drives the number of secondary multiplied charges produced per each primary electron/hole pair. From the combination of such parameters derives one the most critical figure of merit of RSD devices: the readout characteristic time. It has to be sufficiently long to guarantee the capacitive coupling of signals before being discharged through the resistive sheet, but also short enough to minimize pile-up effects. Then the discharge occurs through proper DC contacts, located at the device periphery.

\section{Sensors design and characterization}

In 2019, both float zone and epitaxial $p$-type 6$^{\prime \prime}$ wafers have been processed at FBK through the step-and-repeat (stepper) lithography technique. The aim of the RSD1 batch was to investigate the optimal parameters providing a working AC-LGAD detector suitable for the 4D particle tracking. The electrical tests we performed revealed that the batch was characterized by a uniform process, not only within each wafer but also among the homologous wafers~\cite{2019Mandurrino_34thRD50}. The extensive dynamic characterizations that have been performed, moreover, allowed to fully characterize the sensors under the signal formation and propagation standpoint. In particular, TCT measurements, where a pulsed laser stimulated the secondary charge production through impact ionization, have been extensively used to provide 2D maps of charges induced on the AC-pads~\cite{2019Mandurrino_EDL}. From these studies it was possible to point out that: (\textit{i}) signals are bipolar (due to the capacitive nature of the RSD readout scheme), (\textit{ii}) the charge, in general, is always induced in more than one pad and (\textit{iii}) the amplitude of signals decreases when the laser moves away from a pad, while the peak shifts in time. As a consequence, the integrated charge calculated on the first lobe of the waveform as a function of the distance from the pad center follows typical trends, that we used to model the physical laws driving the signal propagation and attenuation~\cite{2021Tornago_NIMA,2021Siviero_JINST}.

\begin{figure}[!h]\begin{center}
\includegraphics[width=\columnwidth]{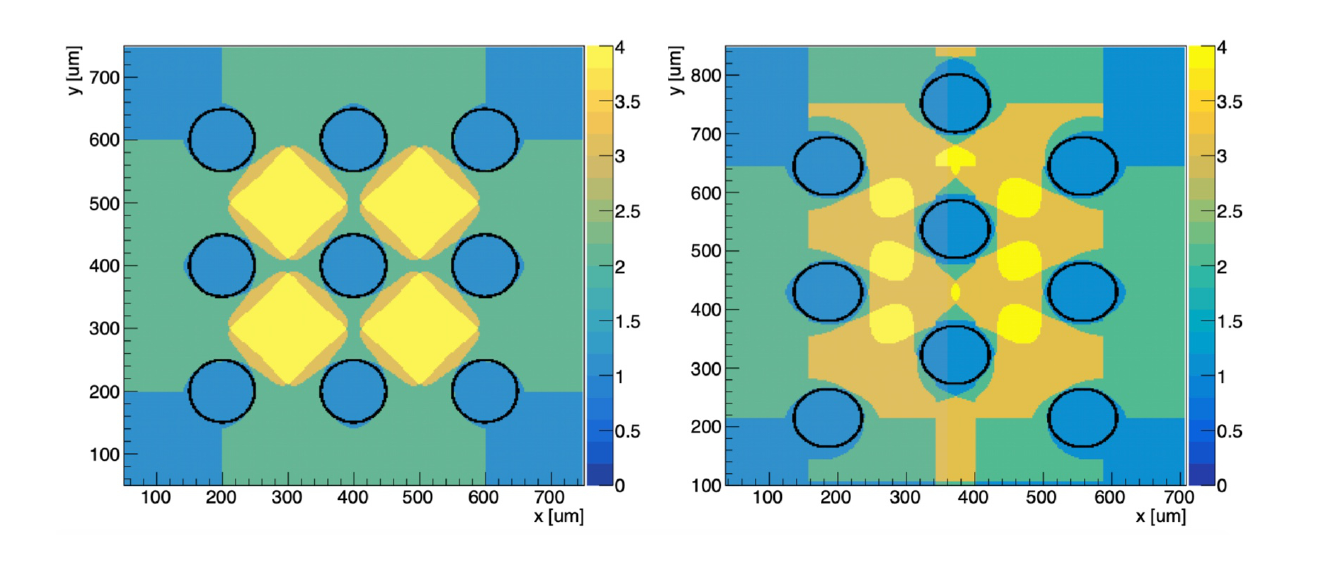}
\caption{\label{fig:ML} Signal distribution among the AC-pads: traditional RSD1 pad array (left) and staggered pad configuration (right). The color scale identifies the number of pads among which the signal is shared, from light blue (1 pad) to yellow (4 pads).}
\end{center}\end{figure}

All the computer calculations we used to predict the dynamic properties of our detectors indicated that in RSD1 the signal sharing, an important figure-of-merit at the basis of the track reconstruction process, involves just one or two pads at the sensor periphery and mostly three or four pads in the inner part of the active region (see the left picture in Figure~\ref{fig:ML}). This means that a beneficial effect such the signal sharing, is likely to become a detrimental effect in terms of track reconstruction if the sensor response is too inhomogeneous. Preliminary machine learning (ML) studies were able to suggest that slight variations of the detector geometry would be sufficient to optimize the signal sharing properties (see the right picture in Figure~\ref{fig:ML}). This observation is at the basis of our second production, RSD2. At any rate, both laser and beam test measurements have shown that RSD can track particles with an unprecedented spatial precision, for an LGAD-based detector, and with a time resolution comparable with traditional trackers with the same gain implant: as an example, a 200-$\mu$m-pitch RSD1 has a space resolution of $\sim$5~$\mu$m and a time resolution of $\sim$40~ps~\cite{2021Tornago_NIMA}.

In early 2021 we launched a second production of RSD. From its release, in September 2021, the RSD2 have been already characterized by the electrical standpoint. No significative variations in breakdown and radiation tolerance properties are expected since the process behind this second production is very similar to that of the first one. Again, we processed fifteen 6$^{\prime \prime}$ wafers (both epitaxial and float zone substrates, respectively 45- and 55-$\mu$m-thick) and all static tests demonstrated that also RSD2 is characterized by a high production quality. The major step forward we implemented in order to push the RSD technology to higher performances in particle tracking lies in the detectors geometry.

\begin{figure}[!h]\begin{center}
\includegraphics[width=\columnwidth]{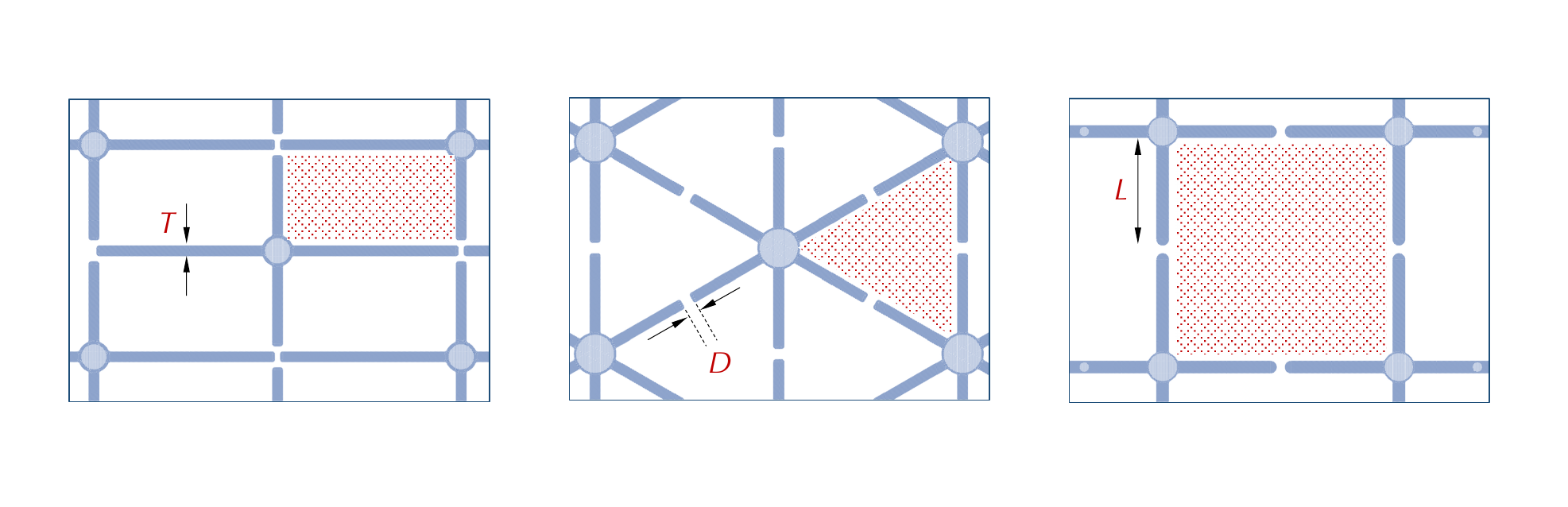}
\caption{\label{fig:frames} AC-pad geometries designed in the RSD2 batch. From left to right: staggered cross-pads with asymmetrical arms, staggered star-pads and regular square array with cross-pads. $T$, and $L$ indicate the thickness and length, respectively, of pad arms while $D$ is the interpad distance.}
\end{center}\end{figure}

If ML algorithms were used just to have a prompt feedback on the role of some macroscopic parameters, like the pad configuration, the TCAD approach was the main strategy we used to determine the details of the final RSD2 layout. As depicted in Figure~\ref{fig:frames}, new pad configurations have been included in the new run~\cite{2021Mandurrino_39thRD50}. In order to obtain the most homogeneous response throughout the sensor active area, staggered geometries, where each pad is located at the corners of an equilateral triangle, have been designed (see the left and middle picture of Figure~\ref{fig:frames}). Also, the inclusion of arms starting from the pad center have been accounted. The arms are intended to produce a confined signal allowing to have always the same number of pads sharing the signals, which is 3 in case of staggered pads or 4 when using the regular square array. When the configuration is triangular and the arms are orthogonal, the only way to confine the signal is by using asymmetrical structures. In all the other cases (triangular arrays with radial arms or regular square arrays) the metal has been shaped in a symmetrical way.

Concerning the parameters $T$, $D$ and $L$, they have also been tuned to test their influence on signal formation and propagation. Typically, $T$ goes from 10 to 20~$\mu$m, while $L$ and $D$ depend on the sensor pitch (50 to 500 $\mu$m) and they are designed to have an interpad distance which is spanning between 50 and 95\% of the pad pitch. This range is so wide because it is crucial to test the influence of different arm lengths on the pad capacitance and, in turn, on the signal shape. Since it was not possible to simulate all the combinations of these parameters, they have been included to be measured afterwards.

\begin{figure}[!h]\begin{center}
\includegraphics[width=\columnwidth]{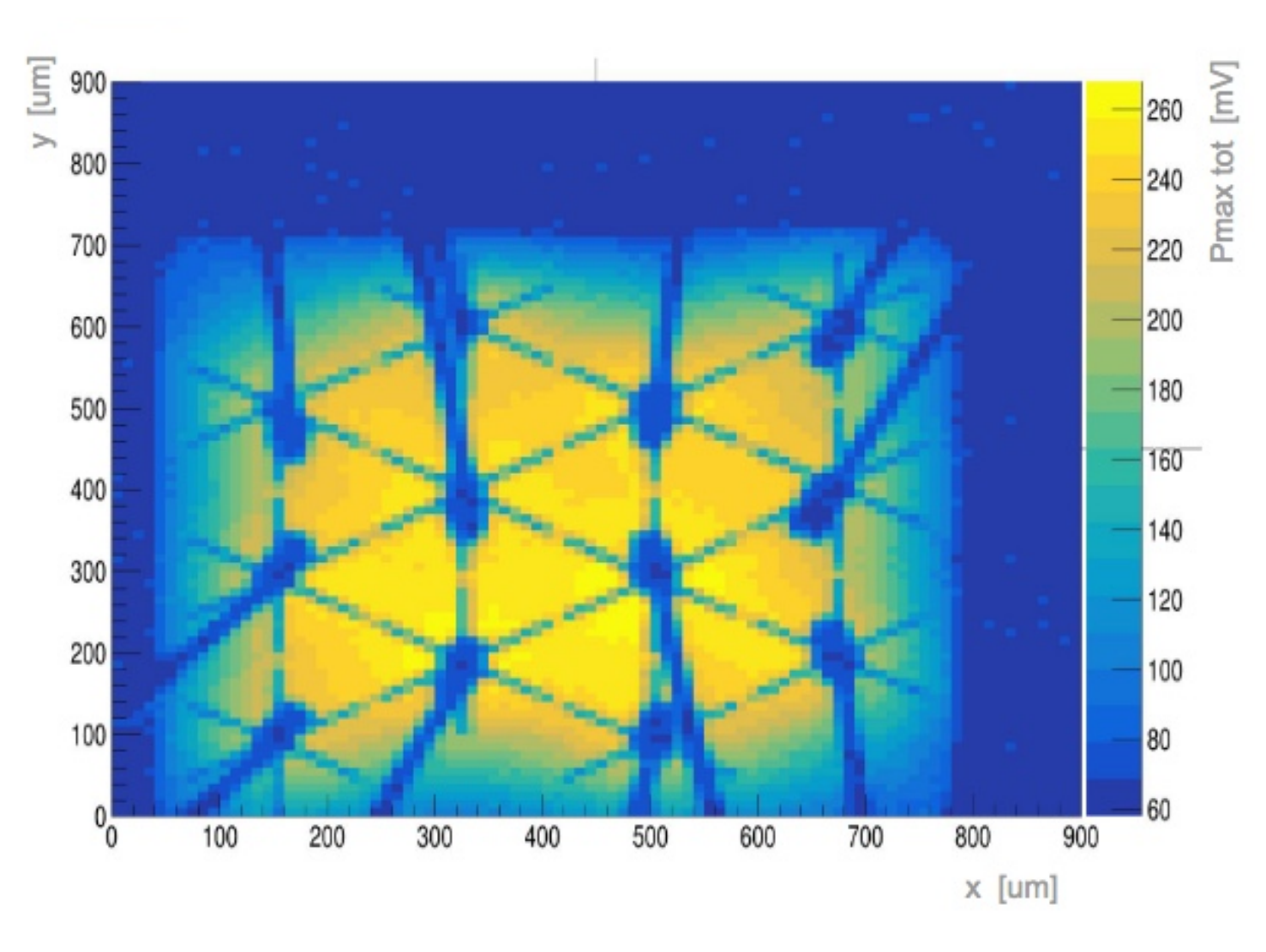}
\caption{\label{fig:RSD2_map} Intensity color map representing the sum of signal amplitudes of all the channels in a 200-$\mu$m-pitch RSD2 with a 3-by-4 pad array in staggered configuration, scanned with an infrared TCT laser.}
\end{center}\end{figure}

As the TCT scan of Figure~\ref{fig:RSD2_map} is indicating, first preliminary laser measurements show that the new pad geometries are able to provide an homogeneous response of the sensors. Here we tested a 200-$\mu$m-pitch RSD2 device were the pads are in staggered configuration and metal arms form triangular boxes. As the signal amplitude acquired at gain 10 suggests, these regions are characterized by uniform response. A feature common for all the detector area (except for a surrounding frame at the periphery, were the triangles are incomplete). Moreover, the signals within each triangular box are always shared by the three first-neighbor pads, as we theoretically expected from the new layout. This is also confirmed by the change of colors occurring beyond the external pads (the above mentioned peripheral frame).

\section*{Acknowledgements}

We kindly acknowledge the following funding agencies, collaborations: INFN, Gruppo V (RSD Project), FBK-INFN collaboration framework, Horizon 2020, Ministero della Ricerca, Italia, PRIN 2017 (progetto 2017L2XKTJ - 4DinSiDe), FARE (R165xr8frt\_fare), Dipartimenti di Eccellenza, Univ. of Torino (ex L. 232/2016, art. 1, cc. 314, 337).

%\section*{References} % comment this line for submission


\begin{thebibliography}{10}

\bibitem{2019Mandurrino_EDL}
M. Mandurrino \emph{et al.}, \emph{Demonstration of 200-, 100-, and 50-$\mu$m pitch Resistive AC-Coupled Silicon Detectors (RSD) with 100\% fill-factor for 4D particle tracking}, \emph{IEEE Electron Device Lett.} {\bf 40}(11) (2019) pp.1780-1783. DOI: https://doi.org/10.1109/LED.2019.2943242

\bibitem{2019Mandurrino_34thRD50}
M. Mandurrino \emph{et al.}, \emph{First production of Resistive AC-Coupled Silicon Detectors (RSD) at FBK}, 34th RD50 Workshop, Lancaster (UK), June 2019. https://indico.cern.ch/event/812761

\bibitem{2021Tornago_NIMA}
M. Tornago \emph{et al.}, \emph{Resistive AC-Coupled Silicon Detectors principles of operation and first results from a combined laser-beam test analysis}, \emph{Nucl. Inst. Meth. A} {\bf 1003} (2021) pg.165319. DOI: https://doi.org/10.1016/j.nima.2021.165319

\bibitem{2021Siviero_JINST}
F. Siviero \emph{et al.}, \emph{First application of machine learning algorithms to the position reconstruction in Resistive Silicon Detector}, \emph{J. Inst.} {\bf 16}(3) (2021) pg.P03019. DOI: https://doi.org/10.1088/1748-0221/16/03/P03019

\bibitem{2021Mandurrino_39thRD50}
M. Mandurrino \emph{et al.}, \emph{RSD2, the new production of AC-LGADs at FBK}, 39th RD50 Workshop, Val\`{e}ncia (Spain), November 2021. https://indico.cern.ch/event/1074989

\end{thebibliography}
\end{document}